\documentclass[11pt,english]{article}
\usepackage[T1]{fontenc}
\usepackage[latin9]{inputenc}
\usepackage{geometry}
\usepackage{mathtools}
\usepackage{amsmath}
\usepackage{amssymb}
\usepackage{graphicx}
\usepackage{setspace}
\makeatletter
\newcommand{\lyxaddress}[1]{
\par {\raggedright #1
\vspace{1.4em}
\noindent\par}
}

\usepackage[numbers,sort&compress]{natbib}

\let\originalleft\left
\let\originalright\right
\renewcommand{\left}{\mathopen{}\mathclose\bgroup\originalleft}
\renewcommand{\right}{\aftergroup\egroup\originalright}

\@ifundefined{showcaptionsetup}{}{%
 \PassOptionsToPackage{caption=false}{subfig}}
\usepackage{subfig}

\makeatother

\usepackage{babel}
\begin{document}

\title{Measurement and Quantum Dynamics in the Minimal Modal Interpretation
of Quantum Theory\thanks{An updated version of this article can be found at\protect\\ \protect\phantom{x* ,} https://drive.google.com/file/d/14fjMeIW-u3byqO9GrCZtqxvH7s-hl2nk/view?usp=sharing}}

\author{Jacob A. Barandes$^{1,}$\thanks{barandes@physics.harvard.edu} $\ $and
David Kagan$^{2,}$\thanks{dkagan@umassd.edu}}
\maketitle

\lyxaddress{\begin{center}
\emph{\small{}$^{1}$Jefferson Physical Laboratory, Harvard University,
Cambridge, MA 02138}\\
\emph{\small{}$^{2}$Department of Physics, University of Massachusetts
Dartmouth, North Dartmouth, MA 02747}
\par\end{center}}
\begin{abstract}
Any realist interpretation of quantum theory must grapple with the
measurement problem and the status of state-vector collapse. In a
no-collapse approach, measurement is typically modeled as a dynamical
process involving decoherence. We describe how the minimal modal interpretation
closes a gap in this dynamical description, leading to a complete
and consistent resolution to the measurement problem and an effective
form of state collapse. Our interpretation also provides insight into
the indivisible nature of measurement\textemdash the fact that you
can't stop a measurement part-way through and uncover the underlying
`ontic' dynamics of the system in question. Having discussed the hidden
dynamics of a system's ontic state during measurement, we turn to
more general forms of open-system dynamics and explore the extent
to which the details of the underlying ontic behavior of a system
can be described. We construct a space of ontic trajectories and describe
obstructions to defining a probability measure on this space.
\end{abstract}

\section{Introduction}

Consider the axioms governing the dynamics of quantum systems as set
out by Dirac and von Neumann \cite{Dirac,vonNeumann}:
\begin{enumerate}
\item \textbf{Unitary Evolution: }When a quantum system is closed, its state
vector evolves according to the Schr{\"o}dinger equation.
\item \textbf{Collapse: }When a measurement is performed, the state vector
collapses into one of the measurement's mutually exclusive outcomes. 
\end{enumerate}
The reference to measurements within these axioms is deeply problematic.
One could posit that measurement is somehow fundamental, but then
how does one rigorously determine in a practical, physical scenario
under what precise circumstances one should declare that a measurement
has taken place? Is there a sharp way to define which kinds of systems
are capable of carrying out measurements and which are not? And if
measurements are \emph{not} fundamental, but are instead processes
carried out on quantum states by measurement devices that register
their results in terms of quantum states of their \emph{own}, how
does one avoid charges of circular reasoning that stem from the assertion
that quantum states are nothing more than collections of probabilities
for measurement outcomes?

Any interpretation of quantum theory that claims to refrain from making
any metaphysical commitments about the state of existence of physical
systems immediately runs into the problem of \emph{requiring} the
existence of systems that can carry out measurements in line with
the standard axioms. On the other had, any \emph{realist} interpretation
of quantum theory that involves fundamental collapse must still grapple
with these issues.\footnote{The Ghirardi-Rimini-Weber interpretation is an example of such an
approach \cite{GRW}.}

An alternative is to formulate a \emph{no-collapse }realist interpretation
of quantum theory. One primary task of such an interpretation is to
explain the \emph{appearance} of collapse. The de Broglie-Bohm pilot-wave
interpretation \cite{Bohm1,Bohm2} and the Everett-DeWitt many-worlds
interpretation \cite{DeWitt,Everett} are both prominent examples
of the no-collapse approach. The traditional pilot-wave interpretation
asserts that particles have simultaneously well-defined positions
and momenta at all times\textemdash these quantities are elements
of the universe's ontology, meaning its fundamental state of being.
The wave function plays the role of a `pilot wave': a field defined
on the system's configuration space that ensures that the particles'
behaviors are effectively captured by the usual rules of quantum theory.
For example, the pilot wave ensures that the measured values of the
particles' momenta exhibit the statistical spread required by the
uncertainty principle. In this situation, as Bohm himself originally
explained, decoherence ensures that collapse emerges purely phenomenologically
from treating both the system to be measured and the measurement device
according to the rules of quantum theory. 

The many-worlds interpretation, like the pilot-wave interpretation,
postulates that the entire universe behaves as a closed quantum system
with a `universal state vector.' In the many-worlds interpretation,
the universal state vector is the key element of the ontology. In
this picture, collapse is only apparent because we are subsystems
that are part of the overall state of the universe, and when we entangle
with the outcome states of a measurement apparatus, the different
outcomes and our perceptions of those outcomes lie on separate `branches'
of the universal state vector. These branches, roughly speaking, are
the interpretation's `worlds.'

Both approaches suffer from severe shortcomings, many of which have
been examined in the literature \cite{InterpretationCritiquesAlbert,InterpretationCritiquesWallace}.
In this paper, we focus on the measurement problem through the lens
of the minimal modal interpretation.\footnote{An in-depth discussion of various types of modal interpretations and
their shortcomings is covered in \cite{Vermaas}.} Unlike the de Broglie-Bohm interpretation, which defines a system's
ontology as always consisting of hidden positions and momenta that
are nowhere to be found within the standard quantum formalism, our
interpretation embraces minimalism by adhering more closely to the
ingredients of textbook quantum theory, allowing the ontology of a
system to evolve dynamically so that the fundamental properties posessed
by the system may change with time. This minimalist approach helps
ensure that our interpretation extends to \emph{all }types of quantum
systems, both relativistic and non-relativistic, with no need for
modification. Furthermore, the minimal modal interpretation is a `single-world'
interpretation in the sense of asserting definite outcomes for measurements
and thereby dodging several of the problems that many-worlds-type
approaches run into. We describe how our interpretation avoids these
problems in \cite{BigMMIpaper}.

In Section 2, we lay out the axioms that define the minimal modal
interpretation. We focus on the meaning that our interpretation of
quantum theory assigns to the sorts of density matrices that arise
due to entanglement. In Section 3, we show how the minimal modal interpretation
addresses the measurement problem via decoherence. Section 4 explores
the extent to which the evolution of the actual state of a quantum
system can be made explicit during processes such as measurements,
as well as the role of non-probabilistic (or `Knightian' \cite{Knightian,Aaronson})
uncertainty in hiding the trajectory of this underlying state. We
conclude in Section 5 with a summary and some ideas for future directions.

\section{The Minimal Modal Interpretation}

\subsection{Axioms}

Historically, modal interpretations have represented a way to talk
about quantum states of systems in terms of \emph{possible} or \emph{actual}
ontic states, where `ontic' denotes an ontological or existential
feature of the universe rather than an aspect of observation. The
minimal modal interpretation identifies a system's density-matrix
eigenstates with the system's set of possible ontic states at any
given moment. The corresponding eigenvalues encode probabilistic uncertainty
about which of these ontic states is actually occupied by the system.
The usual quantum laws governing the time evolution of the density
matrix are supplemented with a set of quantum conditional probabilities
that govern the dynamics of the system's possible ontic states over
time.

More precisely, the interpretation is defined by the following axioms.
\begin{enumerate}
\item \textbf{Ontic States\label{enu:Axiom1}: }A given quantum system at
any particular instant of time $t$ has a mutually exclusive set of
\emph{possible} ontic states $\left\{ \Psi_{i}(t)\right\} $ that
collectively form the system's dynamical configuration space. The
system's \emph{actual} ontic state is one of these possibilities.
The possible ontic states correspond to a set of mutually orthogonal,
unit-norm state vectors in the system's Hilbert space $\mathcal{H}$:
\begin{align}
\Psi_{i}(t) & \iff|\Psi_{i}(t)\rangle\in\mathcal{H}\ \textrm{(defined up to an overall phase),}\\
\Psi_{i}(t),\Psi_{j}(t) & \iff\left\langle \Psi_{i}(t)|\Psi_{j}(t)\right\rangle =\delta_{ij}\ \textrm{(mutual exclusivity)}.
\end{align}
\item \textbf{Objective Epistemic States\label{enu:Axiom2}: }Objective
epistemic states are probability distributions\footnote{We recognize that there are foundational questions about the precise,
rigorous meaning of probability. We wish to disentangle and set aside
these deep mysteries from what we take to be an independent set of
foundational questions in quantum theory. Thus, we merely require
that our probablities obey Kolmogorov's axioms, and we remain agnostic
about the metaphysical meaning of these probabilities. As we shall
see, operationally, the probabilities that define our epistemic states
line up with the empirical outcomes one would expect.} over the system's set of possible ontic states:
\begin{equation}
\left\{ \left(p_{i}(t),\Psi_{i}(t)\right)\right\} _{i},\ \ \ p_{i}(t)\in\left[0,1\right].
\end{equation}
The probabilities here are interpreted as arising from \emph{objective}
uncertainty as to the true ontic state of the system.\footnote{As we will explain later, objective uncertainty can be characterized
as the minimal amount of uncertainty that any observer can attain
regarding the state of the system without perturbing the system. This
kind of uncertainty arises fundamentally from entanglement.} Objective epistemic states can be encoded as density matrices, meaning
positive-semidefinite, unit-trace operators:
\begin{equation}
\left\{ \left(p_{i}(t),\Psi_{i}(t)\right)\right\} _{i}\iff\hat{\rho}\left(t\right)=\sum_{i}p_{i}(t)\left|\Psi_{i}(t)\left\rangle \right\langle \Psi_{i}(t)\right|=\sum_{i}p_{i}(t)\hat{P}\left(i;t\right).
\end{equation}
Here $\hat{P}\left(i;t\right)=\left|\Psi_{i}(t)\left\rangle \right\langle \Psi_{i}(t)\right|$
is the eigenprojector corresponding to the density-matrix eigenstate
$|\Psi_{i}(t)\rangle$ at time $t$. Note, however, that not all density
matrices correspond to objective epistemic states, as we explain in
Section \ref{subsec:The-Interpretation-of}.
\item \textbf{Subsystems and System-Centric Ontology\label{enu:Axiom3}: }
\begin{enumerate}
\item Let $W=Q+E$ denote a parent system consisting of a subsystem $Q$
and its larger environment $E$, and let $\hat{\rho}_{W}$ denote
the density matrix representing the objective epistemic state of $W$.
The objective epistemic state of $Q$ corresponds to the partial trace
\begin{equation}
\hat{\rho}_{Q}(t)\equiv\textrm{Tr}_{E}\left[\hat{\rho}_{W}(t)\right].
\end{equation}
The previous axioms then define the ontic and objective epistemic
states of $Q$.
\item The ontology of a given system is defined \emph{solely} via the eigenvectors
of that system's \emph{own} (reduced) density matrix. The density-matrix
eigenvectors of other systems are irrelevant. We call this notion
\emph{system-centric ontology}.\footnote{The notion of system-centric ontology can also be thought of as a
`localization' of ontology that is reminiscent of the way in which
general relativity localizes inertial reference frames. By contrast,
an interpretation like many-worlds expands the universal state vector
in a preferred basis and defines the ontology of all systems with
respect to that seemingly arbitrary choice. }
\end{enumerate}
\item \textbf{Quantum Conditional Probabilities\label{enu:Axiom4}:} Let
$W=Q_{1}+Q_{2}$ be a parent system that is subdivided into two mutually
disjoint subsystems $Q_{1}$ and $Q_{2}$. Let $\hat{\rho}_{W}\left(t\right)$
be the density matrix describing the parent system's objective epistemic
state at time $t$. When the time evolution of this density matrix
from one time $t$ to another time $t'$ is well-approximated by a
linear completely positive and trace-preserving (CPTP) mapping,
\begin{equation}
\mathcal{E}_{W}^{t'\leftarrow t}:\hat{\rho}_{W}\left(t\right)\mapsto\mathcal{E}_{W}^{t'\leftarrow t}\left\{ \hat{\rho}_{W}\left(t\right)\right\} ,
\end{equation}
 which generalizes the notion of unitary time evolution, we can \emph{define}
quantum conditional probabilities in the following way:
\begin{equation}
p_{Q_{1},Q_{2}|W}\left(i_{1},i_{2};t'|w;t\right)\equiv\textrm{Tr}_{W}\left[\left(\hat{P}_{Q_{1}}\left(i_{1};t'\right)\otimes\hat{P}_{Q_{2}}\left(i_{2};t'\right)\right)\mathcal{E}_{W}^{t'\leftarrow t}\left\{ \hat{P}_{W}\left(w;t\right)\right\} \right].\label{eq:Axiom4}
\end{equation}
This formula gives the joint conditional probability for finding the
subsystems $Q_{1},Q_{2}$ in their respective ontic states $\Psi_{i_{1}},\Psi_{i_{2}}$
at time $t'$ given that the parent system $W$ was in the ontic state
$\Psi_{w}$ at time $t$.
\end{enumerate}
In Appendix \ref{sec:An-Almost-Derivation}, we motivate the definition
in (\ref{eq:Axiom4}) and explain how it generalizes to joint conditional
probabilities for any number of mutually disjoint subsystems. The
conditional probabilities of Axiom \ref{enu:Axiom4} are only sharply
defined to the extent that a system's evolution during a particular
time interval $t\to t'$ is well-approximated by a linear CPTP mapping
$\mathcal{E}_{W}^{t'\leftarrow t}$. When these conditional probabilities
are not sharply defined, the system continues to have an actual ontic
state, but the uncertainty surrounding the time evolution of that
state is non-probabilistic. We explore some of these issues in more
depth in Section \ref{subsec:Non-linear-Evolution}.

In our analysis of the measurement process, we make use of the special
case in which $Q_{1}=W$ is the entire system and $Q_{2}$ is trivial.
We have
\begin{align}
p\left(w';t'|w;t\right) & =\textrm{Tr}\left[\hat{P}\left(w';t'\right)\mathcal{E}^{t'\leftarrow t}\left\{ \hat{P}\left(w;t\right)\right\} \right],\label{eq:evolutionA}
\end{align}
where we have dropped system labels because now only one system is
being considered. The conditional probabilities (\ref{eq:evolutionA})
describe the stochastic result of the time evolution of the system's
ontic state during a process that is approximated by linear CPTP dynamics.
Exact unitary evolution is a particular, idealized case in which the
system in question is completely isolated\textemdash a situation that
is impossible to achieve in the real world.

One must resist the temptation to identify the smooth evolution of
density-matrix eigenprojectors under a smooth, linear CPTP map with
the trajectory of the system's actual ontic state. This trajectory
is generically hidden due to entanglement, as discussed in Section
\ref{sec:OnticDynamics}.

\subsection{The Interpretation of Mixed States\label{subsec:The-Interpretation-of}}

Consider a closed quantum system that is fundamentally described by
a state vector $|\Psi_{1}\rangle$ in a two-dimensional Hilbert space.
Suppose, however, that \emph{you} do not know with certainty that
$|\Psi_{1}\rangle$ is the system's state vector. Instead, letting
$|\Psi_{2}\rangle$ denote the orthogonal state vector, unique up
to overall phase, you can capture your ignorance by introducing a
probability distribution $\left\{ p_{1},p_{2}\right\} $ over the
pair of mutually orthogonal state vectors $\left\{ |\Psi_{1}\rangle,|\Psi_{2}\rangle\right\} $.
The density matrix
\begin{equation}
\hat{\rho}=\sum_{i}p_{i}\left|\Psi_{i}\left\rangle \right\langle \Psi_{i}\right|=p_{1}\left|\Psi_{1}\left\rangle \right\langle \Psi_{1}\right|+p_{2}\left|\Psi_{2}\left\rangle \right\langle \Psi_{2}\right|\label{eq:DMProper}
\end{equation}
then represents an example of what's called a \emph{proper} \emph{mixture}
or a \emph{properly} \emph{mixed} \emph{state}. The interpretation
of this density matrix as encoding subjective ignorance regarding
the system's true state is completely standard and widely accepted.

By contrast, consider a parent system $W=A+B$, with $A$ and $B$
disjoint subsystems. Suppose that the parent system is in the pure
state
\[
c_{1}|\Psi_{A,1},\Psi_{B,1}\rangle+c_{2}|\Psi_{A,2},\Psi_{B,2}\rangle,
\]
where $\left|c_{1}\right|^{2}+\left|c_{2}\right|^{2}=1$ and where
the sets $\left\{ |\Psi_{A,\alpha}\rangle\right\} _{\alpha}$ and
$\left\{ |\Psi_{B,\beta}\rangle\right\} _{\beta}$ consist of mutually
orthogonal vectors in the respective Hilbert spaces of the subsystems
$A$ and $B$. We also employ the standard notational convention
\[
|\Psi_{A,\alpha},\Psi_{B,\beta}\rangle\equiv|\Psi_{A,\alpha}\rangle\otimes|\Psi_{B,\beta}\rangle.
\]
Consider the reduced density matrix of subsystem $A$:
\begin{equation}
\hat{\rho}_{A}\equiv\textrm{Tr}_{B}\left[\hat{\rho}\right]=\left|c_{1}\right|^{2}\left|\Psi_{A,1}\left\rangle \right\langle \Psi_{A,1}\right|+\left|c_{2}\right|^{2}\left|\Psi_{A,2}\left\rangle \right\langle \Psi_{A,2}\right|.\label{eq:DMImproper}
\end{equation}
This density matrix represents an example of an \emph{improper} \emph{mixture}.\emph{
}This density matrix (\ref{eq:DMImproper}) is identical in form to
the density matrix of a proper mixture (\ref{eq:DMProper}), but it
arises in a very different context and, \emph{crucially}, the standard
axioms of quantum theory leave its interpretation unclear. The minimal
modal interpretation fills in this gap by identifying the reduced
density-matrix eigenstates, $|\Psi_{A,1}\rangle$ and $|\Psi_{A,2}\rangle$
with the mutually exclusive possible ontic states that the subsystem
$A$ may occupy. Axiom 1 requires that $A$ in\emph{ }fact occupies
one of these two possibilities. Axiom 2 asserts that the eigenvalues
of this density matrix represent probabilities that describe fundamental
probabilistic uncertainty.

The meaning that we assign to improper mixtures is quite natural:
We interpret them as capturing probabilistic uncertainty about the
state of the system in the same manner as do proper mixtures. The
distinction between proper and improper mixtures lies in the origin
of the uncertainty they encode. Proper mixtures involve \emph{subjective}
uncertainty\textemdash different observers may write down different
density matrices depending on the information they have about the
state of the system. By contrast, improper mixtures involve \emph{objective}
uncertainty\textemdash an irreducible level of uncertainty that is
unique to quantum systems. Objective uncertainty arises due to entanglement
between the system and its environment, and serves to mask the actual
ontic state of the system.\footnote{The minimal modal interpretation can thus be thought of as a hidden-variables
interpretation where the actual ontic state of the system plays the
role of a hidden variable.}

\section{The Measurement Problem}

\subsection{Conceptual Overview}

Suppose that a system of interest\textemdash called here the subject
system\textemdash is initially isolated and described by a given pure
state. Measuring the system opens it up to the environment, resulting
in entanglement between the system, the measurement device, and the
overall environment. If the subject system began in a quantum superposition
of states that can be experimentally distinguished by the measurement
device, then the entanglement process leads to an effective loss of
coherence at the level of the subject system, a process known as decoherence.

But the measurement problem doesn't end with decoherence\textemdash because
decoherence yields an improper mixture. There is a gap in going from
this improperly mixed state to a proper mixture that would immediately
lend itself to describing subjective uncertainty over a unique measurement
outcome. Put a bit differently, if measurements induce entanglement
between a subject system and a measurement device or the larger environment,
why do we see a definite outcome? Decoherence alone cannot answer
this question. In the absence of an explicit collapse postulate, this
issue can only be resolved by going beyond the other traditional axioms
of quantum theory\emph{.}

The axioms of the minimal modal interpretation close this gap by identifying
each system's possible ontic states with the eigenstates of its objective
density matrix. Decoherence forces the subject system's possible ontic
states to converge rapidly to the possible outcome states of the given
measurement, differing only by exponentially small corrections arising
from the residual coherence terms. The measurement device's density
matrix undergoes an analogous rapid convergence to corresponding possible
outcome states. Although neither the subject system's density matrix
nor the measurement device's density matrix are exactly diagonal in
the expected basis, they can be made diagonal by exponentially small
redefinitions of the measurement-outcome states. Furthermore, the
actual ontic state of the subject system is axiomatically taken to
be one of these ontic possibilities, and similarly for the measurement
device, meaning that the overall measurement results in a definite
outcome.

One concern that has been raised historically about modal interpretations
surrounds the status of degeneracies in density matrices\textemdash that
is, the case in which multiple eigenprojectors belonging to a single
density matrix share precisely the same eigenvalue \cite{InterpretationCritiquesAlbert,Vermaas}.
In this scenario, a density matrix does not single out a unique set
of orthonormal eigenstates, leading to ambiguities in the presumed
ontology for the corresponding quantum system.

However, as explained in greater detail in \cite{BigMMIpaper}, true
degeneracies are impossible to realize in practice, as they correspond
to measure-zero arrangements requiring infinite fine-tuning. What
represents a more serious potential issue is the case of \emph{near}-degeneracies
that can arise during the decoherence process, as these near-degeneracies
can ostensibly result in instabilities in the ontic states of \emph{macroscopic}
systems, such as measurement devices and the larger environment. The
minimal modal interpretation's fourth axiom\textemdash defining quantum
conditional probabilities\textemdash ameliorates this issue. As long
as the evolution of the systems in question is linear CPTP to a good
approximation, the conditional probabilities smooth out the evolution
of the ontic states, thereby avoiding undesirable ontic instabilities
\cite{BigMMIpaper}.

\subsection{Measurements and Decoherence\label{subsec:Measurements-and-Decoherence}}

Consider an initially isolated subject system $S$ and a system observable
with orthonormal eigenbasis $\left\{ |m\rangle\right\} _{m}$. A macroscopic
measurement apparatus $A$ is initially prepared in a ``blank''
pure state $|A\left(``\emptyset"\right)\rangle$. The larger environment
$E$ is initially in the pure state $|E\left(``\emptyset"\right)\rangle$,
registering the apparatus's blank state. We assume that the evolution
of the total system is linear CPTP (if not unitary) and can be modeled
as consisting largely of two steps.
\begin{enumerate}
\item If the subject system is in a pure state corresponding precisely to
one of the measurement outcomes $|m\rangle$, then the apparatus transitions
to a new pure state $|A\left(``m"\right)\rangle$ that registers the
state of the subject system:
\begin{equation}
|m\rangle|A\left(``\emptyset"\right)\rangle|E\left(``\emptyset"\right)\rangle\stackrel{\textrm{step }1}{\mapsto}|m\rangle|A\left(``m"\right)\rangle|E\left(``\emptyset"\right)\rangle.
\end{equation}
\item The environment then transitions to a pure state $|E\left(``m"\right)\rangle$
that observably registers in some way the change in the apparatus's
state\textemdash for example, through the transmission of outgoing
thermal radiation:
\begin{equation}
|m\rangle|A\left(``m"\right)\rangle|E\left(``\emptyset"\right)\rangle\stackrel{\textrm{step }2}{\mapsto}|m\rangle|A\left(``m"\right)\rangle|E\left(``m"\right)\rangle.
\end{equation}
\end{enumerate}
Generalizing, suppose now that $|\Psi\rangle$ is the initial pure
state of the subject system. Expanding it in terms of the measurement-outcome
states,
\begin{equation}
|\Psi\rangle=\sum_{m}\Psi_{m}|m\rangle,
\end{equation}
the assumed linearity of the dynamics then dictates that the foregoing
two-step sequence applies to each individual term in the superposition:
\begin{align}
|\Psi\rangle|A\left(``\emptyset"\right)\rangle|E\left(``\emptyset"\right)\rangle & =\left(\sum_{m}\Psi_{m}|m\rangle\right)|A\left(``\emptyset"\right)\rangle|E\left(``\emptyset"\right)\rangle\nonumber \\
 & \stackrel{\textrm{step }1}{\mapsto}\left(\sum_{m}\Psi_{m}|m\rangle|A\left(``m"\right)\rangle\right)|E\left(``\emptyset"\right)\rangle\\
 & \stackrel{\textrm{step }2}{\mapsto}\sum_{m}\Psi_{m}|m\rangle|A\left(``m"\right)\rangle|E\left(``m"\right)\rangle.
\end{align}
After the measurement, the reduced density matrix of the subject system
takes the form
\begin{align}
\hat{\rho}_{S}\left(\Delta t\right) & =\sum_{\mathclap{m_{1},m_{2}}}\Psi_{m_{1}}\Psi_{m_{2}}^{*}\left|m_{1}\left\rangle \right\langle m_{2}\right|\left\langle A\left(``m_{2}"\right)|A\left(``m_{1}"\right)\right\rangle \left\langle E\left(``m_{2}"\right)|E\left(``m_{1}"\right)\right\rangle \nonumber \\
 & =\sum_{m}\left|\Psi_{m}\right|^{2}\left|m\left\rangle \right\langle m\right|+\sum_{\mathclap{\substack{m_{1},m_{2}\\
m_{1}\neq m_{2}
}
}}\Psi_{m_{1}}\Psi_{m_{2}}^{*}\left|m_{1}\left\rangle \right\langle m_{2}\right|\left\langle A\left(``m_{2}"\right)|A\left(``m_{1}"\right)\right\rangle \left\langle E\left(``m_{2}"\right)|E\left(``m_{1}"\right)\right\rangle ,\label{eq:postSDM}
\end{align}
where $\Delta t$ is the time duration of the measurement. The quantities
$\left\langle A\left(``m_{2}"\right)|A\left(``m_{1}"\right)\right\rangle $
and $\left\langle E\left(``m_{2}"\right)|E\left(``m_{1}"\right)\right\rangle $
may not be \emph{exactly }zero for $m_{1}\neq m_{2}$, but they are
exponentially small, as we will now explain. Let $a_{1},a_{2},\ldots,a_{N_{A}}$
be the degrees of freedom of the apparatus and let $e_{1},e_{2},\ldots,e_{N_{E}}$
be the degrees of freedom of the environment, where $N_{A}$ and $N_{E}$
are respectively the numbers of degrees of freedom making up the apparatus
and the environment. Suppose for simplicity that the respective measurement-outcome
states of the apparatus and environment can be factorized in terms
of their degrees of freedom as
\begin{align*}
|A\left(``m"\right)\rangle & =|a_{1}\left(``m"\right)\rangle|a_{2}\left(``m"\right)\rangle\cdots|a_{N_{A}}\left(``m"\right)\rangle,\\
|E\left(``m"\right)\rangle & =|e_{1}\left(``m"\right)\rangle|e_{2}\left(``m"\right)\rangle\cdots|e_{N_{E}}\left(``m"\right)\rangle.
\end{align*}
Then
\begin{align*}
\left\langle A\left(``m_{2}"\right)|A\left(``m_{1}"\right)\right\rangle  & =\left\langle a_{1}\left(``m_{2}"\right)|a_{1}\left(``m_{1}"\right)\right\rangle \cdots\left\langle a_{N_{A}}\left(``m_{2}"\right)|a_{N_{A}}\left(``m_{1}"\right)\right\rangle \sim\exp\left(-N_{A}\times f_{A}\left(\gamma_{A}\Delta t\right)\right),\\
\left\langle E\left(``m_{2}"\right)|E\left(``m_{1}"\right)\right\rangle  & =\left\langle e_{1}\left(``m_{2}"\right)|e_{1}\left(``m_{1}"\right)\right\rangle \cdots\left\langle e_{N_{E}}\left(``m_{2}"\right)|e_{N_{E}}\left(``m_{1}"\right)\right\rangle \sim\exp\left(-N_{E}\times f_{E}\left(\gamma_{E}\Delta t\right)\right),
\end{align*}
where $\gamma_{A}$ and $\gamma_{E}$ are the interaction rates for
each degree of freedom, and where the functions $f_{A}$ and $f_{E}$
are increasing functions of their arguments at least during the duration
of the measurement interval.\footnote{The detailed behavior of these functions depends on how one models
the coupling between the systems. Explicit realizations have been
studied in \cite{Zurek82}.} For simplicity, we assume that these functions are linear to leading
order and then we define $\gamma\equiv\left(N_{A}\gamma_{A}+N_{E}\gamma_{E}\right)/N$,
where $N\equiv N_{A}+N_{E}$. Expressing the subject system's reduced
density matrix (\ref{eq:postSDM}) directly in terms of its post-measurement
eigenstates,
\begin{equation}
\hat{\rho}_{S}\left(\Delta t\right)=\sum_{s}p_{s}(\Delta t)\left|s;\Delta t\left\rangle \right\langle s;\Delta t\right|,
\end{equation}
we expand each member of the subject system's post-measurement ontic
basis in the measurement basis,
\begin{align}
|s;\Delta t\rangle & =\sum_{m}\psi_{s,m}\left(\Delta t\right)|m\rangle\nonumber \\
 & =\psi_{s,m_{s}}\left(\Delta t\right)|m_{s}\rangle+O\left(e^{-N\gamma\Delta t}\right)\times\sum_{\mathclap{\substack{m\\
m\neq m_{s}
}
}}\sigma_{s,m}\left(\Delta t\right)|m\rangle,\ \ \ \ \ \left|\psi_{s,m_{s}}\left(\Delta t\right)\right|\sim1-O\left(e^{-N\gamma\Delta t}\right),\label{eq:sbasis}
\end{align}
where the second equality captures the exponential suppression of
all but one of the measurement-basis states. Therefore,
\begin{align}
\hat{\rho}_{S}\left(\Delta t\right) & =\sum_{s}\Bigg[p_{s}\left(\Delta t\right)\left|\psi_{s,m_{s}}\left(\Delta t\right)\right|^{2}\left|m_{s}\left\rangle \right\langle m_{s}\right|\nonumber \\
 & \qquad\qquad+O\left(e^{-N\gamma\Delta t}\right)\times p_{s}\left(\Delta t\right)\sum_{\mathclap{\substack{m\\
m\neq m_{s}
}
}}\psi_{s,m_{s}}\left(\Delta t\right)\sigma_{s,m}^{*}\left(\Delta t\right)\left|m_{s}\left\rangle \right\langle m\right|\nonumber \\
 & \qquad\qquad+O\left(e^{-N\gamma\Delta t}\right)\times p_{s}\left(\Delta t\right)\sum_{\mathclap{\substack{m\\
m\neq m_{s}
}
}}\sigma_{s,m}\left(\Delta t\right)\psi_{s,m_{s}}^{*}\left(\Delta t\right)\left|m\left\rangle \right\langle m_{s}\right|\nonumber \\
 & \qquad\qquad+O\left(e^{-2N\gamma\Delta t}\right)\times\sum_{\mathclap{\substack{m,m'\\
m,m'\neq m_{s}
}
}}\sigma_{s,m}\left(\Delta t\right)\sigma_{s,m'}^{*}\left(\Delta t\right)\left|m\left\rangle \right\langle m'\right|\Bigg],\label{eq:subDMpostMeasure}
\end{align}
which implies that $p_{s}\left(\Delta t\right)\approx\left|\Psi_{m_{s}}\right|^{2}$,
up to exponentially small corrections. The Born rule associated with
the standard collapse postulate assigns precisely these outcome probabilities,
neglecting the corrections.

As discussed in Section \ref{subsec:The-Interpretation-of}, the results
of decoherence leave us with a post-measurement improper\emph{ }density
matrix that is very close to diagonal in the measurement observable's
eigenbasis. The minimal modal interpretation's first and second axioms
identify the eigenbasis of the subject system's density matrix with
the possible ontic states of the subject system and stipulate, furthermore
that the subject system occupies one of those possible ontic states
in reality. Thus, after the measurement, the subject system inhabits
an ontic state $|s;\Delta t\rangle$ that is exponentially close to
one of the measurement-outcome eigenstates $|m_{s}\rangle$. As we
have seen, the corresponding outcome probabilities are almost precisely
those given by the Born rule, up to exponentially decaying corrections.

\subsubsection*{Decoherence and Quantum Conditional Probabilities}

Introductions to quantum theory often describe a Born probability
as being ``the probability of measuring an outcome state $|m\rangle$
given that the system was in the initial state $|\Psi\rangle$.''
This way of thinking about Born probabilities makes intuitive sense
but can lead to misconceptions due to the lack of a canonical definition
of conditional probabilities in the traditional formulation of quantum
theory. The quantum conditional probabilities defined in Axiom \ref{enu:Axiom4}
allow us to put the foregoing understanding of Born probabilities
on firmer footing.

To this end, suppose that the initial pure state of the subject system
is indeed $|\Psi\rangle$. As we have already seen, each of the system's
possible ontic states $|s;\Delta t\rangle$ at the end of the measurement
process is exponentially close to a corresponding measurement-outcome
state $|m_{s}\rangle$. According to the minimal modal interpretation,
the conditional probability for the system to be in $|s;\Delta t\rangle$
given its initial state $|\Psi\rangle$ is given by
\begin{equation}
p\left(s;\Delta t|\Psi;0\right)=\textrm{Tr}_{S}\left[\hat{P}\left(s;\Delta t\right)\mathcal{E}_{S}^{\Delta t}\left\{ \left|\Psi\left\rangle \right\langle \Psi\right|\right\} \right]=\textrm{Tr}_{S}\left[\hat{P}\left(s;\Delta t\right)\hat{\rho}_{S}\left(\Delta t\right)\right]=\left\langle s;\Delta t\left|\hat{\rho}_{S}\left(\Delta t\right)\right|s;\Delta t\right\rangle .
\end{equation}
Substituting (\ref{eq:postSDM}) yields
\[
p\left(s;\Delta t|\Psi;0\right)=\sum_{\mathclap{m_{1},m_{2}}}\Psi_{m_{1}}\Psi_{m_{2}}^{*}\left\langle s;\Delta t|m_{1}\right\rangle \left\langle m_{2}|s;\Delta t\right\rangle \left\langle A\left(``m_{2}"\right)|A\left(``m_{1}"\right)\right\rangle \left\langle E\left(``m_{2}"\right)|E\left(``m_{1}"\right)\right\rangle .
\]
Equation (\ref{eq:sbasis}) implies that
\begin{equation}
\left\langle s;\Delta t|m_{s'}\right\rangle =\delta_{ss'}+\rho_{ss'}O\left(e^{-N\gamma\Delta t}\right)
\end{equation}
for some coefficients $\rho_{ss'}$ of order 1. After some additional
substitutions, we arrive at
\begin{equation}
p\left(s;\Delta t|\Psi;0\right)=\left|\Psi_{m_{s}}\right|^{2}+O\left(e^{-N\gamma\Delta t}\right).
\end{equation}
Thus, we see that our quantum conditional probabilities reproduce
the correct Born-rule probabilities, again up to exponentially suppressed
corrections.

\subsection{Corrections to the Born Rule and Error-Entropy Bounds}

The minimal modal interpretation and decoherence provide a dynamical
underpinning for the Born rule, but also yield deviations from the
textbook version of the rule. In principle, these deviations show
up in all statistical quantities derived from the traditional Born
rule\textemdash including in all expectation values, final-outcome
states, semiclassical observables, scattering cross sections, tunneling
probabilities, and decay rates. Nevertheless, ordinary measurements
involving macroscopic devices have enormous numbers of degrees of
freedom, rendering the empirical observation of such deviations impractical
in typical circumstances.\footnote{Deviations from the Born rule arise when measurements are modeled
as a decoherence-type quantum process involving a measurement apparatus
and environment of finite size and an interaction of finite duration.
Such deviations are not a unique feature of the minimal modal interpretation,
but also occur in other interpretive frameworks, like the many-worlds
interpretation, in which decoherence plays a central role.}

Setting aside practical considerations, it is interesting to explore
the implications of these exponentially small corrections to the Born
rule. Consider a measurement apparatus $A$ consisting of $N$ degrees
of freedom. The number of degrees of freedom is proportional to the
maximum amount of the system's Shannon (or Gibbs) entropy $S_{\textrm{max}}$,
corresponding to an epistemic state that assigns equal probabilities
to all the possible ontic states of the system.

When the apparatus is used to make a measurement, the interaction
alters the apparatus's density matrix so as to encode a probability
distribution $\left\{ p_{A}\left(``m"\right)\right\} $ that parallels
that of the outcome states for the subject system, as ensured by the
Schmidt decomposition theorem. This probability distribution is associated
with a correlational entropy produced by the measurement according
to
\begin{equation}
S=-\sum_{m}p_{A}\left(``m"\right)\log p_{A}\left(``m"\right).
\end{equation}
The accuracy of the apparatus is bounded from above by the number
of possible ontic states it has available to correlate with the subject
system's states. We can therefore estimate the minimum error of the
apparatus to be the inverse of the number of its possible ontic states.
For our present example, the number of possible ontic states for our
apparatus is exponential $\sim\#^{N}$ in the number $N$ of its degrees
of freedom, and so the minimum error is of order $\#^{-N}\sim e^{-S_{\textrm{max}}}$.

The corrections to the Born rule arising from decoherence are in keeping
with this error-entropy bound, a result completely absent from traditional
or instrumentalist approaches to quantum theory that take the exact
Born rule as an axiom and \emph{derive} the partial-trace from this
and related axioms. In the minimal modal interpretation, by contrast,
the partial-trace operation is an \emph{a priori} ingredient that
can be established independently from the Born rule \cite{BigMMIpaper}.

\section{Ontic State Dynamics\label{sec:OnticDynamics}}

\subsection{Conceptual Overview}

Our discussion of measurement processes so far has involved linear
CPTP time development of a system's objective density matrix. According
to the minimal modal interpretation, this linear CPTP time development
can be thought of as an evolution of the system's set of possible
ontic states combined with the evolution of their corresponding probabilities
of their being the system's actual ontic state. Nowhere in this discussion
have we attempted to map out the system's actual\emph{ }ontic \emph{trajectory}
explicitly.

There are reasons to be suspicious of an interpretation of quantum
theory that claims to pinpoint such an actual ontic trajectory. Interpretive
approaches that attempt to do so would be at risk of violating one
or more of the no-go theorems that constrain all interpretations \cite{Bell,CHSH,GHZ,GHZM,Hardy1,Hardy2,Clifton,Myrvold}.
In the minimal modal interpretation, entanglement hides the actual
ontic state of a system, so it would be inconsistent to expect that
a process that entangles a system with others (such as decoherence)
would allow an explicit description of a specific actual ontic trajectory.

Nevertheless, the minimal modal interpretation suggests a natural
analogy with classical systems, allowing us to specify the nature
of ontic trajectories. A bit more concretely, imagine a classical
system with a fixed set of possible states, which we will take to
be discrete for simplicity. The only way to describe nontrivial evolution
of the actual state of this classical system would be to allow it
to jump between the various discrete possibilities. As we shall see,
the minimal modal interpretation carries this picture over to the
quantum case.

However, there is no \emph{a priori} probability measure on the space
of such ontic trajectories, and therefore no general way to link the
actual ontic dynamics to the evolution of the system's density matrix.
Thus, the uncertainty over the actual ontic trajectory during any
sufficiently smooth, non-unitary evolution of the system (whether
it be a measurement or something more general) is generically not
even probabilistic. We discuss this non-probabilistic uncertainty
further in Section \ref{subsec:Perfectly-Linear-Evolution}.

\subsection{Ontic Trajectories\label{subsec:Ontic-Trajectories}}

According to the minimal modal interpretation, a quantum system has
an actual ontic state at any given instant. This actual ontic state
is a member of a mutually exclusive set of possibilities represented
by all the eigenvectors of the system's objective density matrix.
As time passes, the system's actual ontic state can change, following
a trajectory in the system's Hilbert space. Our goal here is to characterize
these quantum ontic trajectories and determine to what extent they
can be explicitly described.

There is an analogy between quantum ontic trajectories and what we
might call their classical counterparts. Consider a discrete classical
system for which we label the finite number of possible states with
integers $i=1,\ldots,N$. Suppose that we pick out $M$ instants $t_{1},\ldots,t_{M}$
over a finite time interval $t\to t'$, with $t_{1}=t$ and $t_{M}=t'$.
We do not necessarily assume that these instants are evenly spaced.
A classical ontic trajectory is then specified by a sequence of index
choices $\left\{ i\right\} _{t\to t'}=\left(i_{1},\ldots,i_{M}\right)$
representing actual ontic states that the system occupies at the corresponding
times. We imagine that between these times, the classical system's
state jumps from one ontic possibility to another.

For example, suppose that we have a two-state system\textemdash an
idealized, possibly biased coin\textemdash that can jump between its
two states (idealized coin flips). Suppose, moreover, that we consider
ontic trajectories described by four instants $t_{1},t_{2},t_{3},t_{4}$.
One particular such trajectory is given by the sequence $\left(1,2,1,2\right)$,
representing a coin starting in state 1 (heads), jumping to state
2 (tails) at time $t_{2}$, jumping back to heads at $t_{3}$, and
finally jumping to tails at $t_{4}$.

Generalizing to an $N$-state classical system and assuming that the
jumps can occur at any instant, the space of classical ontic trajectories
is simply $\mathbb{Z}_{N}\times\mathbb{R}$, where $\mathbb{R}$ is
the real number line corresponding to time. A particular ontic trajectory
is an integer-valued function $I(t)$ assigning an integer from $\left\{ 1,\ldots,N\right\} $
to each instant in time. Note that for a finite-state system, a nontrivial
trajectory function will not be continuous\textemdash it could even
be \emph{nowhere} continuous.

Turning now to the quantum case, recall that the minimal modal interpretation
identifies the set of eigenstates of the density matrix at any given
instant with the set of possible ontic states that the system may
occupy at that instant. For simplicity, we'll focus on a finite-state
quantum system and label the possible ontic states by integers $i=1,\ldots,N$.
At any fixed instant, the system's actual ontic state is completely
specified by giving an integer that identifies the possible ontic
state that is actualized. A sequence of such integers, one integer
for each instant in time, determines an ontic trajectory\textemdash precisely
like in the finite-state classical case.

But quantum theory also allows for nontrivial evolution of the \emph{basis}
of possible ontic states\textemdash the eigenstates of the density
matrix may evolve over time. This evolution of the basis of possible
ontic states means that the nature of the system's actual ontic state
may change with time, too, even if it doesn't jump to a different
ontic possibility.

Putting these two ideas together, we see that a quantum ontic trajectory
is specified by a choice of elements from $\mathbb{Z}_{N}\times\mathbb{R}$
as well as a unitary operator that twists the set of possible ontic
states around in the system's Hilbert space. Note that this unitary
evolution of the basis of possible ontic states is \emph{not} directly
related to the evolution law for the density matrix as a whole unless
the density matrix as a whole is itself evolving unitarily.

It is also crucial to underline that in all but the most restrictive
circumstances\textemdash namely, a closed system undergoing unitary
evolution\textemdash the actual ontic trajectory of a system is largely
unknowable. However, the system's actual ontic trajectory is constrained
by the minimal modal interpretation's quantum conditional probabilities
whenever the system's dynamics allows these conditional probabilities
to be sufficiently well-defined.

\subsubsection*{Example: A Spin-1/2 System}

\begin{figure}
\begin{centering}
\subfloat[\label{fig:BlochSphere}Depiction of the Bloch sphere.]{
\begin{centering}
\includegraphics[scale=0.4]{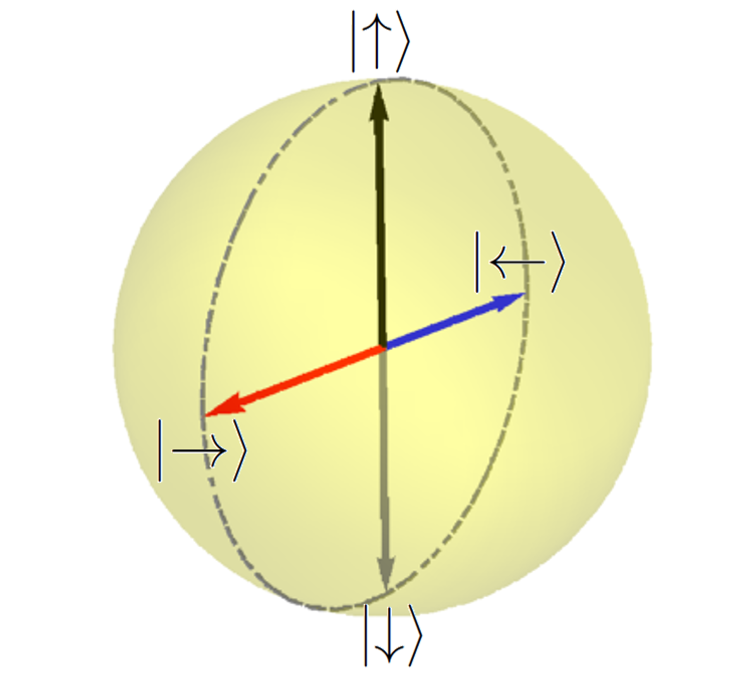}
\par\end{centering}

}\subfloat[\label{fig:BlochSection}Cross-section of the Bloch sphere.]{
\begin{centering}
\includegraphics[scale=0.4]{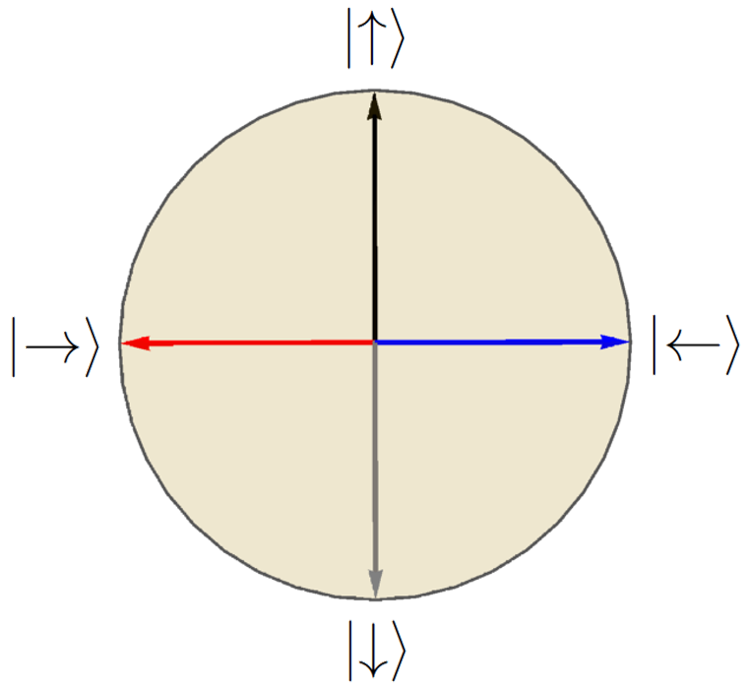}
\par\end{centering}

}\subfloat[\label{fig:ontictraj}An ontic trajectory.]{
\begin{centering}
\includegraphics[scale=0.5]{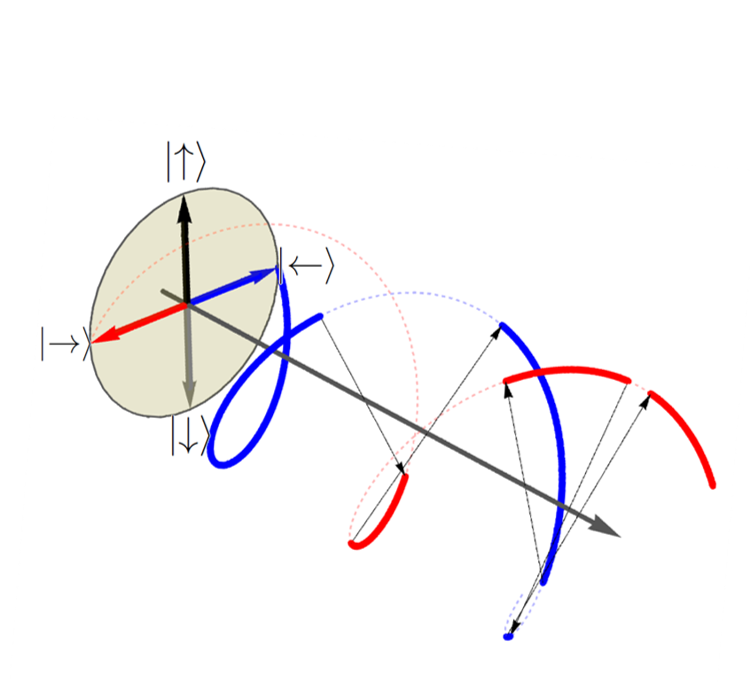}
\par\end{centering}

}
\par\end{centering}
\caption{}

\end{figure}

A spin-1/2 system has a two-dimensional Hilbert space whose rays can
be visualized as points on the surface of a sphere of unit radius,
known as the Bloch sphere (see Figure \ref{fig:BlochSphere}), according
to the following parametrization: 
\begin{equation}
|\Psi\rangle=\cos\frac{\theta}{2}|\uparrow\rangle+e^{i\phi}\sin\frac{\theta}{2}|\downarrow\rangle,\qquad0\leq\theta\leq\pi,\quad0\leq\phi<2\pi.
\end{equation}
This expression is the most general formula for a unit-norm vector
in a two-dimensional complex Hilbert space, up to irrelevant overall
phase. Antipodal points represent orthogonal states\textemdash states
of opposite spin.

Consider the circular slice of the Bloch sphere depicted in Figure
\ref{fig:BlochSection}, with spin right/left along the $x$-axis
and spin up/down along the $z$-axis. Let the unitary evolution of
the ontic basis smoothly and steadily rotate an axis through the Bloch
sphere that is initially aligned with the $x$-axis. We can visualize
this set-up in terms of a cylinder $S^{1}\times\mathbb{R}$, where
$\mathbb{R}$ refers to time, and mark the points at antipodal ends
as they trace out a double helix\textemdash the shape of the trajectories
of the possible ontic states of the system. These helical trajectories
are depicted as red and blue dotted curves in Figure \ref{fig:ontictraj}.

The actual ontic trajectory can hop between the two strands of this
double helix. There can be hops at every instant in time (so the trajectory
can be nowhere continuous), but, for visualization purposes, imagine
that it spends finite time intervals on alternating strands. One such
trajectory is depicted as solid colored red and blue curves segments
in Figure \ref{fig:ontictraj}. The jumps are depicted as thin black
arrows crossing from one possible ontic trajectory to the other.

\subsection{Unitary Evolution and Closed-System Ontic Dynamics}

The dynamics of a system's density matrix is linked to the dynamics
of the underlying ontic state of the system through our quantum conditional
probabilities (\ref{eq:Axiom4}), which again are well-defined to
the extent that the system's overall evolution is at least approximately
linear CPTP. These conditional probabilities constrain the ontic dynamics.
The most straightforward example of such a constraint arises in the
familiar, idealized case of a perfectly closed system whose exact
evolution is given by a unitary operator $\hat{U}\left(t',t\right)$.
In the usual Schr{\"o}dinger picture, the state of the system is given
by a time-dependent, normalized vector in the system's Hilbert space
such that
\begin{equation}
|\Psi\left(t'\right)\rangle=\hat{U}\left(t',t\right)|\Psi\left(t\right)\rangle.
\end{equation}
At any choice of $t'$, we can extend the state vector $|\Psi\left(t'\right)\rangle$
to a (non-unique) set of orthogonal states $\left\{ |\Psi_{k}\left(t'\right)\rangle\right\} _{k}$
that collectively correspond to the set of eigenstates of a degenerate
density matrix, with $|\Psi_{1}\left(t'\right)\rangle\equiv|\Psi\left(t'\right)\rangle$.
In this case, the first eigenvalue of this density matrix is 1 and
the rest vanish, representing the fact that there is no objective
uncertainty as to the state of the system.

We have thus already described the ontic dynamics of this system\textemdash namely,
it never jumps to any other possible ontic state. The quantum conditional
probabilities automatically reflect this condition:
\begin{align*}
p\left(\Psi;t'|\Psi;t\right) & =\textrm{Tr}\left[\left|\Psi\left(t'\right)\left\rangle \right\langle \Psi\left(t'\right)\right|\mathcal{E}^{t'\leftarrow t}\left\{ \left|\Psi\left(t\right)\left\rangle \right\langle \Psi\left(t\right)\right|\right\} \right]\\
 & =\textrm{Tr}\left[\left|\Psi\left(t'\right)\left\rangle \right\langle \Psi\left(t'\right)\right|\hat{U}\left(t',t\right)\left|\Psi\left(t\right)\left\rangle \right\langle \Psi\left(t\right)\right|\hat{U}^{\dagger}\left(t',t\right)\right]\\
 & =1.
\end{align*}
By a similar quick calculation, we find $p\left(\Psi_{k\neq1};t'|\Psi;t\right)=0$.

The time interval in question can be made as small or large as desired
given the idealized assumptions about the system being closed and
unitarily evolving, so the only ontic trajectory available in this
case is the one that tracks the state vector $|\Psi\left(t'\right)\rangle$
in the system's Hilbert space. This unitary case is the unique scenario
for which there is no uncertainty as to the ontic trajectory of the
system.

\subsection{Perfectly Linear CPTP Evolution\label{subsec:Perfectly-Linear-Evolution}}

In order to examine the more general case of linear CPTP evolution,
we begin by considering a closed and therefore unitarily evolving
parent system $W=S+E$ that we decompose into a subject system $S$
and an environment $E$. We can always decompose the density matrix
of $W$ in the following manner:
\begin{equation}
\hat{\rho}_{W}=\hat{\rho}_{S}\otimes\hat{\rho}_{E}+\hat{\sigma}_{SE}.\label{eq:DMwhole-parts}
\end{equation}
Here the respective density matrices $\hat{\rho}_{S}$ and $\hat{\rho}_{E}$
for the two subsytems $S$ and $E$ are defined as usual by the appropriate
partial traces
\begin{equation}
\hat{\rho}_{S}=\textrm{Tr}_{E}\left[\hat{\rho}_{W}\right],\quad\hat{\rho}_{E}=\textrm{Tr}_{S}\left[\hat{\rho}_{W}\right],
\end{equation}
and the correlation operator $\hat{\sigma}_{SE}$, which we can take
to be \emph{defined }by $\hat{\rho}_{W}-\hat{\rho}_{S}\otimes\hat{\rho}_{E}$,
has vanishing partial traces over $S$ and $E$ by construction and
captures the correlations, both classical and quantum, between the
systems $S$ and $E$ \cite{NonlinearEvo1,NonlinearEvo2}. For the
present discussion, we assume that there was an instant at which there
were no correlations between the subsystems\footnote{We will address the more general situation in which correlations always
exist in Section \ref{subsec:Non-linear-Evolution}.} and set that as the initial time $t=0$:
\begin{equation}
\hat{\rho}_{W}\left(0\right)=\hat{\rho}_{S}\left(0\right)\otimes\hat{\rho}_{E}\left(0\right).
\end{equation}
Given such an initially uncorrelated parent density matrix, the unitary
evolution of the parent system reduces to linear CPTP evolution of
the subsystems. In particular, the linear CPTP evolution of $S$ is
given by the map
\begin{equation}
\mathcal{E}_{S}^{t\leftarrow0}\left\{ \hat{\rho}_{S}\left(0\right)\right\} =\textrm{Tr}_{E}\left[\hat{U}_{W}\left(t\right)\hat{\rho}_{S}\left(0\right)\otimes\hat{\rho}_{E}\left(0\right)\hat{U}_{W}^{\dagger}\left(t\right)\right]=\sum_{\mathclap{e,e'}}\hat{E}_{ee'}\left(t\right)\hat{\rho}_{S}\left(0\right)\hat{E}_{ee'}^{\dagger}\left(t\right),\label{eq:LinearEvolution}
\end{equation}
where $\hat{U}_{W}\left(t\right)\equiv\hat{U}_{W}\left(t,0\right)$
is the unitary evolution operator for the parent system $W$. The
summation appearing on the right-hand side of (\ref{eq:LinearEvolution})
is known as the \emph{Kraus} \emph{decomposition} of $\mathcal{E}_{S}^{t\leftarrow0}$,
and the \emph{Kraus} \emph{operators} $\hat{E}_{ee'}\left(t\right)$
are defined by
\begin{equation}
\hat{E}_{ee'}\left(t\right)\equiv\sqrt{p_{e}\left(0\right)}\left\langle e';0\left|\hat{U}_{W}\left(t\right)\right|e;0\right\rangle ,\label{eq:EeeOps}
\end{equation}
where $\left\{ |e;t\rangle\right\} _{e}$ is the system $E$'s set
of density-matrix eigenstates at time $t$. 

We arrive at some critical observations.
\begin{enumerate}
\item[(a)] The linear CPTP evolution map $\mathcal{E}_{S}^{t\leftarrow0}$ for
the subject system $S$ depends on the inital density matrix of the
environment $E$: 
\begin{equation}
\hat{\rho}_{E}\left(0\right)=\sum_{e}p_{e}\left(0\right)\left|e;0\left\rangle \right\langle e;0\right|.
\end{equation}
The definition of $\hat{E}_{ee'}\left(t\right)$ in (\ref{eq:EeeOps})
makes explicit this dependence on the eigenvalues $\left\{ p_{e}\left(0\right)\right\} _{e}$
and on the eigenstates $\left\{ |e;0\rangle\right\} _{e}$.
\item[(b)] The evolution map $\mathcal{E}_{S}^{t\leftarrow0}$ does not in general
respect time-translation symmetry. The initial time $t=0$ is special
because it is at that particular time that we know that $S$ and $E$
are uncorrelated. In fact, continuous time-translation symmetry would
imply that no correlations ever build up between the subsystems and
therefore that the unitary evolution of the parent system $W$ factorizes.
\item[(c)] The evolution map $\mathcal{E}_{S}^{t\leftarrow0}$ is not reversible
in general in the sense of having a well-defined inverse. Nor does
there generically exist a linear CPTP evolution map between two arbitrary
times $t$ and $t'$. The assumed unitarity of the parent system $W$
implies that another linear CPTP evolution map $\mathcal{E}_{S}^{-t\leftarrow0}$
can be defined to evolve backward in time from the uncorrelated state
at $t=0$. However, this evolution map likewise cannot generally be
reversed.
\end{enumerate}
These observations also apply to the quantum conditional probabilities
that constrain the trajectory of the actual ontic state. These conditional
probabilities (a) depend on the initial density matrix of the environment,
(b) are not time-translation invariant, and (c) are only generally
well-defined for intervals either backward or forward from the initial
time $t=0$ when the systems $S$ and $E$ are uncorrelated.

The smoothing out of the actual ontic state's trajectory over short
time intervals provides a concrete illustration of the way that our
quantum conditional probabilities constrain such trajectories. Let
the subsystem $S$ have the initial density matrix
\begin{equation}
\hat{\rho}_{S}\left(0\right)=\sum_{s}p_{s}\left(0\right)\hat{P}\left(s;0\right),
\end{equation}
where each eigenprojector of the initial density matrix is $\hat{P}\left(s;0\right)=\left|s;0\left\rangle \right\langle s;0\right|$.
Similarly, at time $t$,
\begin{equation}
\hat{\rho}_{S}\left(t\right)=\sum_{s'}p_{s'}\left(t\right)\hat{P}\left(s';t\right),
\end{equation}
where each eigenprojector of the density matrix at time $t$ is $\hat{P}\left(s';t\right)=\left|s';t\left\rangle \right\langle s';t\right|$.
Then the quantum conditional probability that the subject system is
in the ontic state $|s';t\rangle$ at time $t$ given an initial ontic
state $|s;0\rangle$ at time $t=0$ is
\begin{align}
p\left(s';t|s;0\right) & =\textrm{Tr}_{S}\left[\hat{P}\left(s';t\right)\mathcal{E}_{S}^{t\leftarrow0}\left\{ \hat{P}\left(s;0\right)\right\} \right]\nonumber \\
 & =\sum_{\mathclap{e,e'}}\textrm{Tr}_{S}\left[\hat{P}\left(s';t\right)\hat{E}_{ee'}\left(t\right)\hat{P}\left(s;0\right)\hat{E}_{ee'}^{\dagger}\left(t\right)\right].\label{eq:QCPlinearCPTP}
\end{align}
In the limit $t\to0$, we have $\hat{E}_{ee'}\left(t\right)\to\sqrt{p_{e}\left(0\right)}\delta_{ee'}\hat{1}$.
Substituting this limiting expression into (\ref{eq:QCPlinearCPTP})
and carrying out the partial trace over $S$ yields the trivial result
\begin{equation}
\lim_{t\to0}p\left(s';t|s;0\right)=\sum_{\mathclap{e,e'}}p_{e}\left(0\right)\delta_{ee'}\delta_{ss'}=\delta_{ss'}.
\end{equation}
Assuming that the parent system's evolution is smooth, then when the
time $t$ is very close to $t=0$, the conditional probability that
the system's actual ontic state jumps between orthogonal possible
ontic states is very small.

Absent any further structure at the level of the parent system's evolution
map, no other quantum conditional probabilities can be defined for
the subsystem $S$ that are conditioned solely on information in $S$,
and no additional constraints on the actual ontic trajectory of $S$
may be inferred. For example, even if we have a definition of the
space of ontic trajectories for the subsystem $S$, the conditional
probabilities defined by (\ref{eq:QCPlinearCPTP}) generally do not
tell us much about which particular trajectory the system's actual
ontic state may follow to some final state $|s';t\rangle$ at time
$t$ given an initial state $|s;0\rangle$. Moreover, one cannot generally
subdivide the time interval from $0$ to $t$ into smaller sub-intervals
and compute intermediate conditional probabilities in the hopes of
establishing some kind of probabilistic measure on the space of ontic
trajectories.

To see more concretely what goes wrong, we suppose for our present
purposes the existence of a linear CPTP evolution map $\mathcal{E}_{S}^{t_{2}\leftarrow t_{1}}$
from a time $t_{1}>0$ to a later time $t_{2}>t_{1}$ and consider
the composition of the two evolution maps $\mathcal{E}_{S}^{t_{1}\leftarrow0}$
and $\mathcal{E}_{S}^{t_{2}\leftarrow t_{1}}$. Introducing a convenient
multi-index $\alpha\equiv\left(e,e'\right)$, and denoting by $\left\{ \hat{E}_{\alpha}^{t_{2}\leftarrow t_{1}}\right\} _{\alpha}$
the Kraus operators corresponding to the evolution map $\mathcal{E}_{S}^{t_{2}\leftarrow t_{1}}$,
we expand the composition of the two evolution maps in terms of their
respective Kraus decompositions:
\begin{align}
\mathcal{E}_{S}^{t_{2}\leftarrow t_{1}}\left\{ \mathcal{E}_{S}^{t_{1}\leftarrow0}\left\{ \hat{\rho}_{S}\left(0\right)\right\} \right\}  & =\sum_{\alpha}\mathcal{E}_{S}^{t_{2}\leftarrow t_{1}}\left\{ \hat{E}_{\alpha}^{t_{1}\leftarrow0}\hat{\rho}_{S}\left(0\right)\left(\hat{E}_{\alpha}^{t_{1}\leftarrow0}\right)^{\dagger}\right\} \nonumber \\
 & =\sum_{\mathclap{\alpha,\beta}}\hat{E}_{\beta}^{t_{2}\leftarrow t_{1}}\hat{E}_{\alpha}^{t_{1}\leftarrow0}\hat{\rho}_{S}\left(0\right)\left(\hat{E}_{\alpha}^{t_{1}\leftarrow0}\right)^{\dagger}\left(\hat{E}_{\alpha}^{t_{2}\leftarrow t_{1}}\right)^{\dagger}.\label{eq:FormalEvo}
\end{align}
Comparing (\ref{eq:FormalEvo}) with the correct evolution according
to the \emph{single} evolution map $\mathcal{E}_{S}^{t_{2}\leftarrow0}$
from $0\to t_{2}$,
\begin{equation}
\mathcal{E}_{S}^{t_{2}\leftarrow0}\left\{ \hat{\rho}_{S}\left(0\right)\right\} =\sum_{\alpha}\hat{E}_{\alpha}^{t_{2}\leftarrow0}\hat{\rho}_{S}\left(0\right)\left(\hat{E}_{\alpha}^{t_{2}\leftarrow0}\right)^{\dagger},
\end{equation}
we note that the Kraus operators in these two expressions do not agree.
This mismatch is a symptom of the fact that composition of linear
CPTP evolution operators does \emph{not} generically obey the semigroup
property
\begin{equation}
\mathcal{E}_{S}^{t_{2}\leftarrow t_{1}}\mathcal{E}_{S}^{t_{1}\leftarrow0}=\mathcal{E}_{S}^{t_{2}\leftarrow0}.\label{eq:LinearComposition}
\end{equation}
The failure of generic linear CPTP evolution to satisfy (\ref{eq:LinearComposition})
means that we cannot safely define a measure on a system's space of
ontic trajectories by subdividing finite time intervals into infinitesimal
pieces and chaining together corresponding conditional probabilities
in a manner analogous to Feynman's formulation of path integrals for
quantum theory.

Thus, in general, even when a parent system has perfectly \emph{unitary
}dynamics, the dynamics of its subsystems can be so unconstrained
that their actual ontic trajectories remain hidden. Subsystem dynamics
cannot even be assumed to provide a probabalistic measure on the space
of ontic trajectories, thereby implying a deeper level of uncertainty
than is usually explicitly considered in physics. Yet, despite its
unfamiliarity, there is nothing illogical about such non-probabilistic
uncertainty. Indeed, it rests on \emph{weaker} assumptions than probabilistic
uncertainty.\footnote{Another area in which such non-probabilistic uncertainty may arise
is cosmology. Models exhibiting eternal inflation generically feature
causally disconnected regions that conceivably manifest different
phases of an underlying physical theory with different empirical properties,
such as different masses for elementary particles and different interaction
couplings between them. There is no obvious way to define a measure
on these empirical attributes. Many attempts have been made and will
likely continue, but the possibility of a more fundamental type of
uncertainty should not be dismissed.}

There is a connection between the hidden nature of the actual ontic
state of an entangled subsystem and the hidden nature of its trajectories.
When a system is entangled with its environment, the minimal modal
interpretation asserts that the system's actual ontic state is one
of the eigenstates of its density matrix (Axioms \ref{enu:Axiom1}
and \ref{enu:Axiom2}). However, the minimal modal interpretation
does not tell us which eigenstate is actualized\textemdash this information
is hidden from all possible observers. For nontrivial linear CPTP
evolution, a system initially uncorrelated with its environment will
begin to build up environmental correlations. Just as entanglement
fundamentally masks the instantaneous actual ontic state of a system,
these evolving correlations mask the details of the system's ontic
trajectory as the system evolves away from its initially uncorrelated
state. We conjecture that our quantum conditional probabilities represent
the maximal amount of information that can be gleaned regarding these
ontic trajectories for general situations governed by a well-defined,
linear CPTP evolution map. Of course, as we have seen, more information
may be available in certain special situations, such as the case of
unitary evolution. We turn now to a more general case in which extra
information regarding ontic trajectories is available.

\subsection{Markovian Evolution}

Consider again the parent system $W=S+E$ of the previous section,
but now suppose that there is a specific time $t_{1}$ at which the
parent system state has `forgotten' the correlations between the subject
system and the environment:
\begin{equation}
\hat{\rho}_{W}\left(t_{1}\right)=\hat{\rho}_{S}\left(t_{1}\right)\otimes\hat{\rho}_{E}\left(t_{1}\right).
\end{equation}
In this case, we could define linear CPTP evolution for the subsystem
$S$ starting at this instant $t_{1}$ in terms of the parent system's
unitary evolution. Self-consistency will then imply the semigroup
property (\ref{eq:LinearComposition}).

Expanding on this observation, we see that if the parent system's
dynamics sets a natural time scale $\delta t_{S}$ beyond which correlations
are periodically mostly erased, then coarse-grained dynamical maps
exist that approximately compose over adjacent time intervals and
are therefore effectively Markovian. As an illustrative example, imagine
a subject system $S$ that is interacting with a bath of photons.
The photons can be thought of as `pinging' against the subject system,
becoming entangled with it, and then rapidly flying off. The local\emph{
}environment $E$ thereby loses its memory of the subject system's
evolution history and the overall state of the parent system $W=S+E$
approximately re-factorizes. This idea has been fruitfully applied
in many areas of physics, and in particular underpins the Lindblad
equation, which provides a differential description of such Markovian
open-system dynamics \cite{Lindblad,Joos,Hornberger}.

The actual ontic trajectory of a system under such circumstances is
still non-probabilistically uncertain for time intervals shorter than
$\delta t_{S}$. But one can define coarse-grained trajectories whose
uncertainty is captured by products of conditional probabilities at
each time-step.

To be explicit, suppose that we have an $N$-state quantum system
for which we consistently enumerate the possible ontic states as 1
through $N$. Let $\Delta t\gg\delta t_{S}$ be the coarse-graining
time interval, let $t_{0}$ be an initial time, and let 
\[
t_{n}=t_{0}+n\Delta t
\]
be a time at which the system's actual ontic state is the possible
ontic state labeled by $i_{n}\in\left\{ 1,\ldots,N\right\} $. We
envision a coarse-grained version of ontic jumps potentially taking
place during the transition times $\Delta t$. In the present context,
there is an approximately well-defined quantum conditional probability
$p\left(i_{n};t_{n}|i_{n-1};t_{n-1}\right)$ for the system to be
in the actual ontic state $i_{n}$ at time $t_{n}$ given that its
actual ontic state was $i_{n-1}$ at time $t_{n-1}$. Given a sequence
of such actual ontic states at particular coarse-grained times $\left\{ \left(i_{n};t_{n}\right)\right\} _{n=0}^{M}$,
the probability associated with this overall coarse-grained ontic
trajectory is 
\begin{equation}
p_{\left\{ i_{k}\right\} }=\prod_{\mathclap{n=1}}^{M}p\left(i_{n};t_{n}|i_{n-1};t_{n-1}\right).\label{eq:CoarseGrainedMeasure}
\end{equation}
This description of coarse-grained ontic trajectories is in keeping
with the literature on quantum trajectories associated with the Lindblad
equation and other related types of open-system dynamics \cite{Hornberger,EspositoMukamel}.

The coarse-grained trajectories described here recapitulate the form
of the exact ontic trajectories of Section \ref{subsec:Ontic-Trajectories}.
The crucial difference, again, is that whereas we have been able to
define a measure (\ref{eq:CoarseGrainedMeasure}) on our \emph{coarse-grained
}trajectories, no measure can be placed \emph{a priori} on the space
of \emph{exact} ontic trajectories during a finite time interval.
Having the additional structure of approximate Markovianity over a
dynamically determined coarse-graining time scale allows for a somewhat
more smoothed-out picture of approximate ontic trajectories.

\subsection{More General Evolution\label{subsec:Non-linear-Evolution}}

It turns out that linear CPTP evolution is not sufficiently general
to capture every possible kind of open-system dynamics. Indeed, open-system
dynamics need not even be linear.\footnote{Abrams and Lloyd argue in \cite{AbramsLloyd} that the freedom to
implement \emph{arbitrarily} \emph{chosen} nonlinear dynamics would
lead to surprising implications for solving NP-complete problems.
We emphasize that the nonlinear dynamics here is \emph{not} fully
under experimental control.}$^{,}$\footnote{ From time to time, one reads of proposals that linear open-system
dynamics can, in fact, be defined even in the presence of initial
subsystem-environment correlations. However, because any such dynamical
map has the specific correlations of a particular initial density
matrix built into its definition, the dynamical map manifestly cannot
be linear in the sense that it can take as inputs general linear combinations
of arbitrary initial density matrices.}

To make this point clear with an explicit example, consider a parent
system $W=S+E$ with subsystems $S$ and $E$ that may never be uncorrelated,
and let $\mathcal{E}_{W}^{t'\leftarrow t}$ be a linear CPTP map for
the parent system $W$, where we leave open the possibility that this
map may be unitary. The density matrix for the subsystem $S$ at time
$t'$ is then given by
\begin{equation}
\hat{\rho}_{S}\left(t'\right)=\textrm{Tr}_{E}\left[\hat{\rho}_{W}\left(t'\right)\right]=\textrm{Tr}_{E}\left[\mathcal{E}_{W}^{t'\leftarrow t}\left\{ \hat{\rho}_{W}\left(t\right)\right\} \right],\label{eq:DMevo}
\end{equation}
which is not generically a linear (or even an analytic) function of
the density matrix of the subsystem $S$ at the initial time $t$,
due to the possible presence of entanglement between the subsystems
$S$ and $E$. Denoting by $\hat{P}_{W}\left(w;t\right)$ each eigenprojector
of $\hat{\rho}_{W}\left(t\right)$ and denoting by $\hat{P}_{S}\left(s';t'\right)$
each eigenprojector of $\hat{\rho}_{S}\left(t'\right)$, we see that
the quantum conditional probabilities constraining ontic trajectories
of $S$ in this more general case explicitly depend on the state of
the parent system $W$:
\begin{equation}
p_{S|W}\left(s';t'|w;t\right)=\textrm{Tr}_{W}\left[\left(\hat{P}_{S}\left(s';t'\right)\otimes\hat{1}_{E}\right)\mathcal{E}_{W}^{t'\leftarrow t}\left\{ \hat{P}_{W}\left(w;t\right)\right\} \right].\label{eq:ConditionalOnWorld}
\end{equation}
Quantum entanglement at the initial time $t$ precludes us from factorizing
the eigenprojectors $\hat{P}_{W}\left(w;t\right)$ of the parent system's
density matrix into a tensor product of eigenprojectors of the respective
density matrices of the subsystems $S$ and $E$.

In the \emph{absence} of entanglement at the initial time $t$, this
factorization can indeed be carried out and we have
\begin{equation}
\hat{P}_{W}\left(w=(s,e);t\right)=\hat{P}_{S}\left(s;t\right)\otimes\hat{P}_{E}\left(e;t\right).
\end{equation}
Then (\ref{eq:ConditionalOnWorld}) becomes
\begin{align}
p_{S|W}\left(s';t'|w;t\right) & =\textrm{Tr}_{W}\left[\left(\hat{P}_{S}\left(s';t'\right)\otimes\hat{1}_{E}\right)\left(\mathcal{E}_{W}^{t'\leftarrow t}\left\{ \hat{P}_{S}\left(s;t\right)\otimes\hat{P}_{E}\left(e;t\right)\right\} \right)\right]\nonumber \\
 & =\textrm{Tr}_{S}\left[\hat{P}\left(s';t'\right)\mathcal{E}_{S|E,e}^{t'\leftarrow t}\left\{ \hat{P}_{S}\left(s;t\right)\right\} \right]\nonumber \\
 & =p_{S|E,e}\left(s';t'|(s,e);t\right),\label{eq:ConditionalJOint}
\end{align}
where
\begin{equation}
\mathcal{E}_{S|E,e}^{t'\leftarrow t}\left\{ \hat{P}_{S}\left(s;t\right)\right\} \equiv\textrm{Tr}_{E}\left[\mathcal{E}_{W}^{t'\leftarrow t}\left\{ \hat{P}_{S}\left(s;t\right)\otimes\hat{P}_{E}\left(e;t\right)\right\} \right]
\end{equation}
is the time-evolution map for the subsystem $S$ conditioned on the
state $e$ of the environment $E$ at time $t$ at which the two subsystems
are not entangled. The conditional-probability formula (\ref{eq:ConditionalJOint})
is analogous to the classical case, for which the probability of the
system occupying a particular ontic state at a later time is conditioned
on the joint state of the two subsystems at an earlier time. By summing
over all environment states $e$, and using the resolution of the
identity $\sum_{e}\hat{P}_{E}\left(e;t\right)=\hat{1}_{E}$, we arrive
at a linear CPTP map $\mathcal{E}_{S}^{t'\leftarrow t}$ describing
the evolution of $S$ without any reference to the environment. So
we see that linear dynamics at the level of a parent system descends
to linear dynamics for a subsystem when that subsystem is not entangled
with other subsystems.

Even in the restricted case in which the parent system's time-evolution
map factorizes,
\begin{equation}
\mathcal{E}_{W}^{t'\leftarrow t}=\mathcal{E}_{S}^{t'\leftarrow t}\otimes\mathcal{E}_{E}^{t'\leftarrow t},\label{eq:FactorizedDynamics}
\end{equation}
the evolution map (\ref{eq:DMevo}) does not generically yield a linear
mapping at the level of the subsystem $S$. Entanglement is therefore
responsible for a uniquely quantum obstruction to the existence of
exactly linear CPTP open-system dynamics. This obstruction represents
a novel, purely quantum form of uncertainty that arises at the level
of ontic dynamics even when the subsystems are not dynamically coupled
in the sense of (\ref{eq:FactorizedDynamics}).

\section{Conclusion}

\subsection{Measurement and Ontic Trajectories}

To what extent can we say anything about the ontic trajectory of a
quantum system during a typical von Neumann-type measurement interaction?
Recalling the simple model of Section \ref{subsec:Measurements-and-Decoherence},
we assume that the subject system is prepared in a state vector $|\Psi\rangle$
and that the measurement apparatus and environment are both initially
uncorrelated with the subject system. We further assume for simplicity
that the parent system evolves unitarily, with decoherence arising
through the entangling of the subject system with its environment.
These features of our simple model imply that we are in the scenario
described by Section \ref{subsec:Perfectly-Linear-Evolution}\textemdash the
subsystem of interest evolves according to a linear CPTP map. We see
that during the measurement's short\textemdash but finite\textemdash span
of time, there are no stringent constraints on the subject system's
underlying ontic dynamics other than that the subject system ends
up in an actual ontic state very close to one of the measurement eigenstates.

Of course, we can cut the measurement process off shortly after it
starts. In that eventuality, our quantum conditional probabilities
dictate that the subject system's actual ontic state will likely still
be quite close to $|\Psi\rangle$. But this fact doesn't tell us anything
about the overall ontic trajectory that the subject system would take
if we were instead to let the measurement develop over a sufficiently
long period.

Clearly, collapse-like dynamics and the Copenhagen interpretation
are successful in practice. From our perspective, this success stems
from the rapid rate of decoherence and the generic indivisibility
of linear evolution that we described earlier. The minimal modal interpretation
threads the needle between, on the one hand, closing the conceptual
gap that decoherence leaves open between proper and improper mixtures
in standard quantum theory, and, on the other hand, saying too much
about the underlying dynamics of ontic states. The role of fundamental,
non-probabilistic uncertainty in masking ontic trajectories is crucial
to the consistency of the interpretation and its minimal modifications
to the standard foundational axioms of quantum theory.

\subsection{Classical vs. Quantum Ontic Trajectories}

The classical and quantum ontic trajectories discussed in Section
\ref{subsec:Ontic-Trajectories} are quite similar. The classical
space of ontic trajectories for an $N$-state system is $\mathbb{Z}_{N}\times\mathbb{R}$,
corresponding to the assignment of a system state to each instant
in time. According to the minimal modal interpretation of quantum
theory, the space of ontic trajectories for an $N$-state quantum
system is essentially the same space of functions as in the classical
case, along with all possible unitary-evolution maps that evolve the
basis of possible ontic states within the quantum system's Hilbert
space.

The foregoing kinematical description alone is insufficient for determining
conditional probabilities over time without additional assumptions
about dynamics. As discussed in Section \ref{subsec:Perfectly-Linear-Evolution},
Markovian behavior isn't generic, even for a quantum system evolving
according to linear CPTP dynamics. A new obstruction to Markovianity
arises from evolving entanglement between the system in question and
its environment. The lack of Markovianity precludes us from defining
a measure on the system's space of ontic trajectories, thereby implying
another, deeper level of non-probabilistic uncertainty about the behavior
of the system's actual ontic state.

In Section \ref{subsec:Non-linear-Evolution}, we explored the scenario
in which a system and its environment may always feature a significant
amount of entanglement. In the presence of such entanglement, the
dynamics a subsystem inherits from its parent system governing the
subsystem's density matrix will be nonlinear. Nonlinear evolution
obstructs even the definition of quantum conditional probabilities,
meaning that the dynamics of the subsystem's actual ontic state is
completely hidden and thus exhibits another manifestation of non-probabilisitc
uncertainty.

\subsection{Does the Minimal Modal Interpretation Modify Quantum Theory?}

We emphasize that the role played by linear CPTP dynamics in the minimal
modal interpretation's axioms is not a fundamental modification of
the dynamics of quantum theory. Our interpretation is formulated in
recognition of the fact that all physically realistic systems are
to some extent open systems and that \emph{standard} quantum theory
exhibits a form of non-reductionism: States of parent systems do not
fully determine the states of their subsystems. In the minimal modal
interpretation, this non-reductionism translates into system-centric
ontology: The ontologies of subsystems do not \emph{a priori }mesh
together in a classically intuitive way that determines the ontology
of the parent system. (See Axiom \ref{enu:Axiom3}.) However, these
system-centric ontologies fit together in a more classically coherent
manner as systems become macroscopic, as we discuss in more depth
in \cite{BigMMIpaper} and will explore further in future work.

\section*{Acknowledgments}

D. K. thanks Gaurav Khanna, Darya Krym, John Estes, and Paul Cadden-Zimansky
for many useful discussions. D. K. has been supported in part by FQXi
minigrant Observers in Quantum Theory-\#10610. J. A. B. would like
to acknowledge helpful conversations with David Albert, Ned Hall,
and Jeremy Butterfield. We are both grateful to Brian Greene and Allan
Blaer for many discussions and insightful suggestions. This is a pre-print of an article published in {\em Foundations of Physics}. The final authenticated version is available online at: https://doi.org/10.1007/s10701-020-00374-0

\appendix

\section{Quantum Conditional Probabilities\label{sec:An-Almost-Derivation}}

In this appendix, we motivate the formula (\ref{eq:Axiom4}) for the
quantum conditional probabilities at the heart of the minimal modal
interpretation. We start with a parent system $W=Q_{1}+Q_{2}$ partitioned
into subsystems $Q_{1}$ and $Q_{2}$ that are mutually disjoint.
The reduced density matrix of the subsystem $Q_{1}$ at time $t'$
is given by the partial trace
\begin{equation}
\hat{\rho}_{Q_{1}}\left(t'\right)=\textrm{Tr}_{Q_{2}}\left[\hat{\rho}_{W}\left(t'\right)\right].
\end{equation}
The reduced density matrix of the subsystem $Q_{2}$ is similarly
defined. At any given time $t$, the density matrices of $W$ and
the subsystems $Q_{1}$ and $Q_{2}$ can be expanded in the bases
of their respective eigenprojectors $\left\{ \hat{P}_{W}\left(w;t\right)\right\} _{w}$,
$\left\{ \hat{P}_{Q_{1}}\left(i_{1};t\right)\right\} _{i_{1}},$ and
$\left\{ \hat{P}_{Q_{2}}\left(i_{2};t\right)\right\} _{i_{2}}$:
\begin{align}
\hat{\rho}_{W}\left(t\right) & =\sum_{w}p_{W}\left(w;t\right)\hat{P}_{W}\left(w;t\right),\\
\hat{\rho}_{Q_{1}}\left(t\right) & =\sum_{i_{1}}p_{Q_{1}}\left(i_{1};t\right)\hat{P}_{Q_{1}}\left(i_{1};t\right),\\
\hat{\rho}_{Q_{2}}\left(t\right) & =\sum_{i_{2}}p_{Q_{2}}\left(i_{2};t\right)\hat{P}_{Q_{2}}\left(i_{2};t\right).
\end{align}

According to the minimal modal interpretation, the probability of
the subsystem $Q_{1}$ being in the ontic state $\Psi_{i_{1}}\left(t'\right)$
at time $t'$ is
\begin{equation}
p_{Q_{1}}\left(i_{1};t'\right)=\textrm{Tr}_{Q_{1}}\left[\hat{P}_{Q_{1}}\left(i_{1};t'\right)\hat{\rho}_{Q_{1}}\left(t'\right)\right].
\end{equation}
This expression can be rewritten as a formula that explicitly involves
the disjoint subsystem $Q_{2}$ and the parent system $W$ by expanding
the trace to encompass the entire parent system's Hilbert space and
inserting an identity operator for the Hilbert space of the subsystem
$Q_{2}$:
\begin{equation}
p_{Q_{1}}\left(i_{1};t'\right)=\textrm{Tr}_{W}\left[\left(\hat{P}_{Q_{1}}\left(i_{1};t'\right)\otimes\hat{1}_{Q_{2}}\right)\hat{\rho}_{W}\left(t'\right)\right].
\end{equation}
The identity operator $\hat{1}_{Q_{2}}$ can be expanded in terms
of the eigenprojectors $\hat{P}_{Q_{2}}\left(i_{2};t'\right)$,
\begin{equation}
\hat{1}_{Q_{2}}=\sum_{i_{2}}\hat{P}_{Q_{2}}\left(i_{2};t'\right),
\end{equation}
and this summation can be pulled out of the trace to yield
\begin{equation}
p_{Q_{1}}\left(i_{1};t'\right)=\sum_{i_{2}}\textrm{Tr}_{W}\left[\left(\hat{P}_{Q_{1}}\left(i_{1};t'\right)\otimes\hat{P}_{Q_{2}}\left(i_{2};t'\right)\right)\hat{\rho}_{W}\left(t'\right)\right].\label{eq:Subsystem1Prob}
\end{equation}
We now suppose that the parent system's evolution is well-approximated
by a linear CPTP map $\mathcal{E}_{W}^{t'\leftarrow t}$. Linearity
implies that
\begin{equation}
\hat{\rho}_{W}\left(t'\right)=\mathcal{E}_{W}^{t'\leftarrow t}\left\{ \hat{\rho}_{W}\left(t\right)\right\} =\sum_{w}p\left(w;t\right)\mathcal{E}_{W}^{t'\leftarrow t}\left\{ \hat{P}_{W}\left(w;t\right)\right\} ,
\end{equation}
thereby allowing us to rewrite (\ref{eq:Subsystem1Prob}) as 
\begin{equation}
p\left(w';t'\right)=\sum_{\mathclap{i_{2},w}}\textrm{Tr}_{W}\left[\left(\hat{P}_{Q_{1}}\left(i_{1};t'\right)\otimes\hat{P}_{Q_{2}}\left(i_{2};t'\right)\right)\mathcal{E}_{W}^{t'\leftarrow t}\left\{ \hat{P}_{W}\left(w;t\right)\right\} \right]p\left(w;t\right).\label{eq:conditionalprob}
\end{equation}
This last expression can be interpreted as a Bayesian propagation
formula in its familiar sense, 
\begin{equation}
p\left(w';t'\right)=\sum_{\mathclap{i_{2},w}}p_{Q_{1},Q_{2}|W}\left(i_{1},i_{2};t'|w;t\right)p\left(w;t\right),
\end{equation}
provided that we adopt Axiom \ref{enu:Axiom4} and make the identification
\begin{equation}
p_{Q_{1},Q_{2}|W}\left(i_{1},i_{2};t'|w;t\right)=\textrm{Tr}_{W}\left[\left(\hat{P}_{Q_{1}}\left(i_{1};t'\right)\otimes\hat{P}_{Q_{2}}\left(i_{2};t'\right)\right)\mathcal{E}_{W}^{t'\leftarrow t}\left\{ \hat{P}_{W}\left(w;t\right)\right\} \right].\label{eq:qcondprob}
\end{equation}
Our last step is not strictly necessary\textemdash we \emph{choose
}to interpret the trace formula in (\ref{eq:qcondprob}) as a conditional
probability. In keeping with the minimalist spirit of our interpretation
of quantum theory, note that we have constructed this new set of conditional
probabilities out of standard ingredients without introducing any
exotic elements or assumptions.

The conditional probabilities defined by (\ref{eq:qcondprob}) can
be generalized to the case of a parent system $W=Q_{1}+\cdots+Q_{n}$
consisting of $n$ disjoint subsystems $Q_{1},\ldots,Q_{n}$ by replacing
$Q_{2}\to Q_{2}+\cdots+Q_{n}$ and $\hat{1}_{Q_{2}}\to\hat{1}_{Q_{2}}\otimes\cdots\otimes\hat{1}_{Q_{n}}$.
Following steps analogous to those detailed above for the bipartite
case, one derives the $n$-subsystem joint conditional probabilities
\begin{equation}
p_{Q_{1},\ldots,Q_{n}|W}\left(i_{1},\text{\ensuremath{\ldots},}i_{n};t'|w;t\right)=\textrm{Tr}_{W}\left[\left(\hat{P}_{Q_{1}}\left(i_{1};t'\right)\otimes\cdots\otimes\hat{P}_{Q_{n}}\left(i_{n};t'\right)\right)\mathcal{E}_{W}^{t'\leftarrow t}\left\{ \hat{P}_{W}\left(w;t\right)\right\} \right].
\end{equation}

Of course, in order for these quantities to qualify as proper conditional
probabilities, they should be non-negative and sum to unity. What
follows is a proof that our quantum conditional probabilities indeed
have these properties.
\begin{enumerate}
\item Non-negativity: The tensor-product operator
\begin{equation}
\hat{P}_{Q_{1}}\left(i_{1};t'\right)\otimes\cdots\otimes\hat{P}_{Q_{n}}\left(i_{n};t'\right)
\end{equation}
and the time-evolved projection operator $\mathcal{E}_{W}^{t'\leftarrow t}\left\{ \hat{P}_{W}\left(w;t\right)\right\} $
are both manifestly positive semidefinite. If we call the first positive
semidefinite operator $\hat{A}$ and the second $\hat{B}$, then $\sqrt{\hat{A}}$
and $\sqrt{\hat{B}}$ are also positive semidefinite and we have
\[
\textrm{Tr}\left[\hat{A}\hat{B}\right]=\textrm{Tr}\left[\sqrt{\hat{A}}\sqrt{\hat{A}}\sqrt{\hat{B}}\sqrt{\hat{B}}\right]=\textrm{Tr}\left[\sqrt{\hat{B}}\sqrt{\hat{A}}\sqrt{\hat{A}}\sqrt{\hat{B}}\right]=\textrm{Tr}\left[\left(\sqrt{\hat{A}}\sqrt{\hat{B}}\right)^{\dagger}\left(\sqrt{\hat{A}}\sqrt{\hat{B}}\right)\right]\geq0.
\]
Therefore, our conditional probabilities are non-negative, as claimed:
\begin{equation}
p_{Q_{1},\ldots,Q_{n}|W}\left(i_{1},\ldots,i_{n};t'|w;t\right)\geq0.
\end{equation}
\item Unit measure: Taking a fixed parent-system ontic state $w$ and summing
over all the final subsystem states $i_{1},\ldots,i_{n}$, we find
\begin{align*}
\sum_{\mathclap{i_{1},\ldots,i_{n}}}p_{Q_{1},\ldots,Q_{n}|W}\left(i_{1},\ldots,i_{n};t'|w;t\right) & =\sum_{\mathclap{i_{1},\ldots,i_{n}}}\textrm{Tr}_{W}\left[\left(\hat{P}_{Q_{1}}\left(i_{1};t'\right)\otimes\cdots\otimes\hat{P}_{Q_{n}}\left(i_{n};t'\right)\right)\mathcal{E}_{W}^{t'\leftarrow t}\left\{ \hat{P}_{W}\left(w;t\right)\right\} \right]\\
 & =\textrm{Tr}_{W}\left[\left(\hat{1}_{Q_{1}}\otimes\cdots\otimes\hat{1}_{Q_{n}}\right)\mathcal{E}_{W}^{t'\leftarrow t}\left\{ \hat{P}_{W}\left(w;t\right)\right\} \right]\\
 & =\textrm{Tr}_{W}\left[\mathcal{E}_{W}^{t'\leftarrow t}\left\{ \hat{P}_{W}\left(w;t\right)\right\} \right]\\
 & =1.
\end{align*}
\end{enumerate}
\noindent 

\end{document}